# Dynamic load-balancing on multi-FPGA systems: a case study

Volodymyr V. Kindratenko
*NCSA, UIUC*
*kindr@ncsa.uiuc.edu*

Robert J. Brunner, Adam D. Myers
*Department of Astronomy, UIUC*
*{rb|adm}@astro.uiuc.edu*

## Abstract

*In this case study, we investigate the impact of workload balance on the performance of multi-FPGA codes. We start with an application in which two distinct kernels run in parallel on two SRC-6 MAP processors. We observe that one of the MAP processors is idle 18% of the time while the other processor is fully utilized. We investigate a task redistribution schema which serializes the execution of the two kernels, yet parallelizes execution of each individual kernel by spreading the workload between two MAP processors. This implementation results in a near 100% utilization of both MAP processors and the overall application performance is improved by 9%.*

## 1. Introduction

Reconfigurable computing (RC) [1] has evolved to the point where it can accelerate computationally intensive, floating-point scientific codes beyond what is possible on conventional microprocessor-based systems [2]. Consequently, in the past few years considerable efforts have been made to port various computational kernels to reconfigurable hardware and to measure and understand their performance. As an example, we have implemented numerous computational kernels on different hardware architectures including the NAMD molecular dynamics code [3], the MATPHOT algorithm [4], and the two-point angular correlation function (TPACF) [5, 6] and, in doing so, we have obtained significant performance improvements in comparison to conventional microprocessor-based platforms. However, in general, fewer attempts have been made to understand the performance of an entire system—one in which coarse-grain functional parallelism can be exploited through conventional parallel processing in addition to the instruction-level parallelism available through direct hardware execution. Yet, this is a very important issue to address, one which can help us to understand the viability of the approach of using heterogeneous microprocessor/reconfigurable processor systems for high-performance computing (HPC) applications.

In this case study, we investigate the issue of workload balancing and its impact on the performance of codes running on multi-FPGA systems. We present an initial parallel implementation of a two-point angular correlation function algorithm on an SRC-6 reconfigurable computer in which the workload is distributed between two microprocessors and four field-programmable gate array (FPGA) chips [6]. The dual-MAP SRC-6 reconfigurable computer used in this study allows us to investigate how a misbalance in the workload can affect the overall performance of the application and it also provides us with a testbed to experiment with various strategies to improve the workload balance on the multi-FPGA systems. We observe that in the initial implementation, one of the MAP processors is idle 18% of the time while the other processor is fully utilized. In order to make better use of the available hardware recourses, we investigate a task redistribution schema that serializes the execution of the two kernels, and parallelizes execution of each individual kernel by spreading the workload between the two MAP processors. This implementation results in a near 100% utilization of both MAP processors, while improving the overall performance by 9%.

The paper is organized as follows: First, we provide an overview of the general problem, including related work, a description of the TPACF algorithm, and the hardware platform used in this study. Section 3 describes our initial dual-MAP implementation of the TPACF algorithm, and continues by detailing our final load-balanced implementation. We conclude the paper with a summary of the lessons learned and a brief discussion.

## 2. Background materials

### 2.1. Related work

Load balancing of tasks executed on FPGA-based accelerator boards has been examined in the context of multiple processing stages executed on the FPGA device itself. Thus, in [7], a dynamic load balancing strategy is proposed and analyzed on an example of a parallel LU factorization of large, sparse block-diagonal-bordered matrices on a configurable multiprocessor. Load balancing is achieved with the help of a dedicated system controller that is aware of the workload on all processors and makes decisions about task distribution based on the processing elements availability. In [8], an on-chip architecture that supports dynamic load balancing for molecular dynamics algorithms is proposed. Load balancing is primarily achieved through the use of specialized processing units that are able to switch, as needed, between various tasks required by the algorithm in order to adapt to the input parameters. In [9], load-balancing of various tasks involved in the discrete elements method calculation engine implemented on an FPGA is achieved via a pre-implementation numerical analysis of the computational requirements for various subroutines employed in the code.

While the existing work is mostly concerned with the problem of load-balancing internal to an FPGA design, in this case study we consider a problem of a dynamic load-balancing of the computations distributed across multiple FPGAs.

## 2.2. Case study problem

As a test case, we consider a problem of computing the two-point angular correlation function, $\omega(\theta)$, as used in the field of Cosmology [10]. A detailed description of the underlying mathematical model used to compute TPACF can be found in [6]; here we provide only a brief summary of the computational kernel.

The computational core of the TPACF algorithm consists of computing the separation distances between the points on a sphere and binning them into distributions at some angular resolution. The binning schema used in this work is logarithmic: each decade of angular distance in the logarithmic space is divided equally between $k$ bins, meaning that there are $k$ equally-logarithmically-spaced bins between, for example, 0.01 and 0.1 arcminutes of angular separation. The problem of computing the angular separation distributions can be expressed as follows:

- **Input:** Set of points $x_1, .., x_n$ distributed on the surface of a sphere, and a small number $M$ of bins: $[\theta_0, \theta_1), [\theta_1, \theta_2), .., [\theta_{M-1}, \theta_M)$.
- **Output:** For each bin, the number of unique pairs of points $(x_i, x_j)$ for which the angular distance is in the respective bin: $B_l = |\{ij: \theta_{l-1} <= x_i \cdot x_j < \theta_l\}|$.

Calculation of the angular distance $\theta$ between a pair points on the sphere requires converting the spherical coordinates to Cartesian coordinates (which is done only once when the data is loaded from the disk), computing their dot product, and taking the arccosine of the computed dot product. Once the angular distance is known, it can be mapped into the respective angular bin $B_l$. A faster approach is to project the bin edges, $\{[\theta_i, \theta_{i+1}); i=0,..,M-1\}$, to the pre-arccosine "dot product" space and to locate the corresponding bin in this space instead of computing the arccosine for each dot product. Since the bin edges are ordered, an efficient binary search algorithm can be used to quickly locate the corresponding bin in just $log_2M$ steps. We therefore adopt this approach to determine the binned counts.

Three types of separation distributions are required for $\omega(\theta)$ to be computed: between all unique pairs of points in the observed dataset (an autocorrelation), between the points from the observed dataset and the points from some number, $n_R$, of random datasets (a cross-correlation), and between unique pairs of points in each of the $n_R$ random data sets (an autocorrelation). Note that formally, the calculation of the cross-correlation requires $N_D^2$ steps whereas the autocorrelation is computed in $(N_D(N_D-1)/2)$ steps.

As a test case dataset, we use a sample of photometrically classified quasars and random catalogs first analyzed by [11] to calculate $\omega(\theta)$. We specifically use one hundred random samples ($n_R=100$); the actual dataset and all of the random realizations each contain 97,178 points. In addition, we employ a binning schema with five bins per decade ($k=5$), $\theta_{min}=0.01$ arcminutes, and $\theta_{max}=10,000$ arcminutes. Thus, angular separations are spread across 6 decades of scale and require 30 bins ($M=30$).

## 2.3. Case study platform

The SRC-6 MAPstation [12] used in this work consists of a commodity, dual-CPU 2.8 GHz Intel Xeon board, one MAP Series C and one MAP Series E processor, and an 8 GB common memory module, all interconnected with a 1.4 GB/s low latency Hi-Bar™ switch. The SNAP™ Series B interface board is used to connect the CPU board to the Hi-Bar switch.

The MAP Series C processor module contains two user FPGAs, one control FPGA, and memory. There are six banks (A-F) of on-board memory (OBM); each bank is 64 bits wide and 4 MB deep for a total of 24 MB. There is an additional 4 MB of dual-ported memory dedicated to data transfer between the two FPGAs. The two user FPGAs in the MAP Series C are Xilinx Virtex-II XC2V6000 FPGAs. The FPGA clock rate of 100 MHz is set from within the SRC programming environment. The MAP Series E processor module is identical to the Series C module with the exception of the user FPGAs: the two user FPGAs in the MAP Series E are Xilinx Virtex-II Pro XC2VP100 chips.

Code for SRC-6 MAPstation is developed in the MAP C programming language using the Carte™ version 2.2 programming environment [13]. The Intel C (icc) version 8.1 compiler is used to generate the CPU-side of the combined CPU/MAP executable. The SRC MAP C compiler produces the hardware description of the FPGA design for our final, combined CPU/MAP target executable. This intermediate hardware description of the FPGA design is passed to Xilinx ISE place and route tools to produce the FPGA bit file. Finally, the linker is invoked to combine the CPU code and the FPGA hardware bit file(s) into a unified executable.

## 3. Case study

The microprocessor-based C implementation of the case study problem is straightforward (Figure 1): pre-compute bin boundaries for a specified range of bins; compute autocorrelation for the observed data; for each random data file, compute autocorrelation and cross-correlation bin counts; and finally use the computed bin counts to calculate the angular correlation function. In this implementation the autocorrelation subroutine is responsible for 33.3% of the execution time whereas the cross-correlation subroutine is responsible for 66.6% of the overall execution time. Pseudo code of the cross-correlation subroutine is shown in Figure 2; the autocorrelation subroutine is similar.

```
(binb, nb) = pre-compute_bin_boundaries();
(d1, n1) = load_observed_data_from_file();
dd = autocorrelation(d1, n1, binb, nb);
for each random data file
   (d2, n2) = load_random_data_from_file();
   rr += autocorrelation(d2, n2, binb, nb);
   dr += cros-scorrelation(d1, n1, d2, n2, binb, nb);
compute_tpacf(dd, rr, dr);
```

**Figure 1**: Pseudo code of the TPACF algorithm.

Since the binary search is invoked after each dot product calculation, performance of the reference C implementation is less dependent on the floating point performance and is bound by the time spent in the binary search. We observe that when executed on the SRC-6 host processor, less than 90 MFLOPS (about 1.5% of peak floating point performance of the processor) is typically achieved.

MAP C implementation of the autocorrelation and cross-correlation subroutines is straightforward (Figure 3): transfer *d1* and *d2* datasets containing *n1* and *n2* points, respectively, from the system memory to the MAP processor OBM banks; loop over all pairs of points in *d1* and *d2*; for each such pair, compute dot product, find the bin it belongs to, and update its count by adding 1; at the end, transfer out a small array of bin counts back to the system memory. The bin finding procedure is implemented using the MAP C *select_pri_8bit_32val* macro, which is an equivalent of a cascaded *if/if else* statement. This macro replaces the most-inner *while* loop in the pseudo code shown in Figure 2 with an unrolled version such that the next most-inner loop, *for (j = 0; j < n2; j++)*, becomes the most-inner loop and can be fully pipelined by the MAP C compiler. As a result, on each iteration of this loop, after some number of initial clock cycles, a new pair of points is processed and a new result is stored.

```
int[] cross-correlation(d1, n1, d2, n2, binb, nbins)
{
    for (i = 0; i < n1; i++)
    {
      for (j = 0; j < n2; j++)
      {
        // compute dot product
        dotp = d1_i.x*d2_j.x + d1_i.y*d2_j.y + d1_i.z*d2_j.z;

        // run binary search
        min = 0, max = nbins;
        while (max > min+1) {
          k = (min + max) / 2;
          if (dot >= binb[k]) max = k;
          else min = k;
        };

        // update bin counts
        bin[max] += 1;
      }
    }
    return bin;
}
```

**Figure 2**: Pseudo code of the cross-correlation kernel.

In practice, we extend the implementation shown in Figure 3 to take the full advantage of the MAP processor resources. Thus, we manually unroll the most inner loop to compute several dot products/bin values simultaneously. The exact number of such simultaneous kernels depends on the size of the FPGA chip used. Table 1 summarizes the number of unrolled steps implement on each of the chips of each of the MAPs available in our system. We use the MAP Series C processor to compute both autocorrelation and cross-correlation and the MAP Series E processor to compute cross-correlation only.

```
#define NBINS 32

void cross_correlationMAPE(double data1[], int n1, double data2[],
int n2, int64_t data_bins[], double binb[], int mapnum)
{
   OBM_BANK_A (AL, double, 262144)
   OBM_BANK_B (BL, double, 262144)
   OBM_BANK_C_2_ARRAYS (CL, double, 262144, CLd, double, 128)
   OBM_BANK_D_2_ARRAYS (DL, double, 262144, DLi, int64_t, 128)
   OBM_BANK_E (EL, double, 262144)
   OBM_BANK_F (FL, double, 262144)

   // bin boundaries
   double bv01, bv02, bv03, bv04, bv05, bv06, bv07, bv08;
   double bv09, bv10, bv11, bv12, bv13, bv14, bv15, bv16;
   double bv17, bv18, bv19, bv20, bv21, bv22, bv23, bv24;
   double bv25, bv26, bv27, bv28, bv29, bv30, bv31;

   // bin counts
   int64_t bin1a[NBINS], bin2a[NBINS], bin3a[NBINS], bin4a[NBINS];

   Stream_64 S0, S1;
   int i, j, bank, indx;
   double dot, pj_x, pj_y, pj_z, pi_x, pi_y, pi_z;

   // load bin boundaries
   #pragma src parallel sections
   {
      #pragma src section
      {
         stream_dma_cpu(&S0, PORT_TO_STREAM, CLd, DMA_C,
                                 binb, 1, (NBINS-1)*8);
      }
      #pragma src section
      {
         for (i = 0; i < NBINS-1; i++) {
            bv01 = bv02; bv02 = bv03; bv03 = bv04; bv04 = bv05;
            bv05 = bv06; bv06 = bv07; bv07 = bv08; bv08 = bv09;
            bv09 = bv10; bv10 = bv11; bv11 = bv12; bv12 = bv13;
            bv13 = bv14; bv14 = bv15; bv15 = bv16; bv16 = bv17;
            bv17 = bv18; bv18 = bv19; bv19 = bv20; bv20 = bv21;
            bv21 = bv22; bv22 = bv23; bv23 = bv24; bv24 = bv25;
            bv25 = bv26;  bv26 = bv27; bv27 = bv28; bv28 = bv29;
            bv29 = bv30; bv30 = bv31;
            get_stream_dbl(&S0, &bv31);
         }
      }
   }

   // DMA data in
   #pragma src parallel sections
   {
      #pragma src section
      {
         // DMA dataset #1 into OBM A-C
         DMA_CPU(CM2OBM, AL, MAP_OBM_stripe(1,"A,B,C"),
                     data1, 1, n1*3*8, 0);
         wait_DMA(0);

         // DMA dataset #2 into OBM D-F
         DMA_CPU(CM2OBM, DL, MAP_OBM_stripe(1,"D,E,F"),
                     data2, 1, n2*3*8, 0);
         wait_DMA(0);
      }
```

```
      #pragma src section
      {
         for (i = 0; i < NBINS; i++)    // reset bin values
         {
            bin1a[i] = 0; bin2a[i] = 0; bin3a[i] = 0; bin4a[i] = 0;
         }
      }
   }

   // main compute loop
   for (i = 0; i < n1; i++) {
      pi_x = AL[i];  pi_y = BL[i];  pi_z = CL[i];   // point i

      #pragma loop noloop_dep
      for (j = 0; j < n2; j++) {
         // what bin memory bank to use in this loop iteration
         cg_count_ceil_32 (1, 0, j == 0, 3, &bank);

         pj_x = DL[j];  pj_y = EL[j];  pj_z = FL[j];    // point j
         dot = pi_x * pj_x + pi_y * pj_y + pi_z * pj_z;  // dot product

        // find what bin it belongs to
         select_pri_64bit_32val( (dot < bv31), 31, (dot < bv30), 30,
           (dot < bv29), 29,(dot < bv28), 28, (dot < bv27), 27,
           (dot < bv26), 26, (dot < bv25), 25, (dot < bv24), 24,
           (dot < bv23), 23, (dot < bv22), 22, (dot < bv21), 21,
           (dot < bv20), 20, (dot < bv19), 19, (dot < bv18), 18,
           (dot < bv17), 17, (dot < bv16), 16, (dot < bv15), 15,
           (dot < bv14), 14, (dot < bv13), 13, (dot < bv12), 12,
           (dot < bv11), 11, (dot < bv10), 10, (dot < bv09), 9,
           (dot < bv08), 8,  (dot < bv07), 7,  (dot < bv06), 6,
           (dot < bv05), 5, (dot < bv04), 4,  (dot < bv03), 3,
           (dot < bv02), 2,  (dot < bv01), 1,  0, &indx);

         // update the corresponding bin count
                 if (bank == 0) bin1a[indx] += 1;
         else if (bank == 1) bin2a[indx] += 1;
         else if (bank == 2) bin3a[indx] += 1;
                        else bin4a[indx] += 1;
      }
   }

   // DMA bins back to the host
   #pragma src parallel sections
   {
      #pragma src section
      {
         int64_t val;
         for (j = 0; j < NBINS; j++) {
            val = bin1a[j] + bin2a[j] + bin3a[j] + bin4a[j];
            put_stream(&S1, val, 1);
         }
      }

      #pragma src section
      {
         stream_dma_cpu(&S1, STREAM_TO_PORT, DLi, DMA_D,
                              data_bins, 1, NBINS*8);
      }
   }
}
```

**Figure 3**: MAP C implementation of the cross-correlation kernel.

We also benefit from the ability to implement an arbitrary precision numerical data type. Thus, when using double-precision floating-point for the bin boundaries, as shown in Figure 3, we can place 2 kernels per chip for the autocorrelation subroutine and 3 kernels per chip for the cross-correlation subroutine (last but one column in Table 1). However, a more detailed numerical analysis shows that a 43-bit, fixed-point data type is sufficient to cover the necessary range of scales used to store the bin boundaries in this particular application. Therefore, when we replace the double-precision floating-point comparison operator with a custom, 43-bit fixed-point comparison operator [14], we can place a larger number of kernels per chip (last column in Table 1). The rest of this paper focuses on this fixed-point implementation.

**Table 1**: Number of compute kernels implemented per FPGA for the double-precision and fixed-point kernels.

| Processor (subroutine) | per FPGA and total | double-precision kernel | fixed-point kernel |
|---|---|---|---|
| MAP Series C (autocorrelation) | primary chip | 2 | 3 |
| | secondary chip | 2 | 4 |
| | Total | 4 | **7** |
| MAP Series E (cross-correlation) | primary chip | 3 | 5 |
| | secondary chip | 3 | 5 |
| | Total | 6 | **10** |

## 3.1. Parallel dual-MAP implementation

The implementation shown in Figure 1 can be trivially parallelized on a multi-processor system. For example, on a two-processor system, such as our MAPstation, the autocorrelation and cross-correlation subroutines can be executed simultaneously, one on each CPU. The same approach applies when porting the subroutines to two MAPs as well, each MAP independently executes one subroutine. In our original implementation [6], we use OpenMP to enable a parallel execution of the MAP-based subroutines (Figure 4).

Figure 5 contains performance results obtained for this straightforward, parallel implementation. We run this code using different sized datasets (horizontal axis in Figure 5) and measure the execution time of the autocorrelation and cross-correlation subroutines (plotted on the left-hand vertical axis). The CPU performance (blue line in Figure 5) is obtained for the implementation shown in Figure 1, and the dual-MAP performance (red line) is obtained for the implementation shown in Figure 4. The speedup is computed as the ratio of the CPU time to the dual-MAP time and is plotted on the right-hand vertical axis in Figure 5 (green bars). As is usually the case, once the dataset size reaches a certain limit, the effects of data transfer overhead become unnoticeable and the overall speedup of the dual-MAP implementation stays roughly constant, around 89x for this application as compared to a single 2.8 GHz Intel Xeon chip.

```
(binb, nb) = pre-compute_bin_boundaries();
(d1, n1) = load_data_from_file();
dd = autocorrelation(d1, n1, binb, nb);
for each random data file
   (d2, n2) = load_data_from_file();
   #pragma omp parallel sections
   #pragma omp section
   rr += autocorrelationMAPC(d2, n2, binb, nb);
   #pragma omp section
   dr += cros-scorrelationMAPE(d1,n1,d2,n2 binb, nb);
compute_w(dd, rr, dr);
```

**Figure 4**: Pseudo code of the parallel TPACF algorithm.

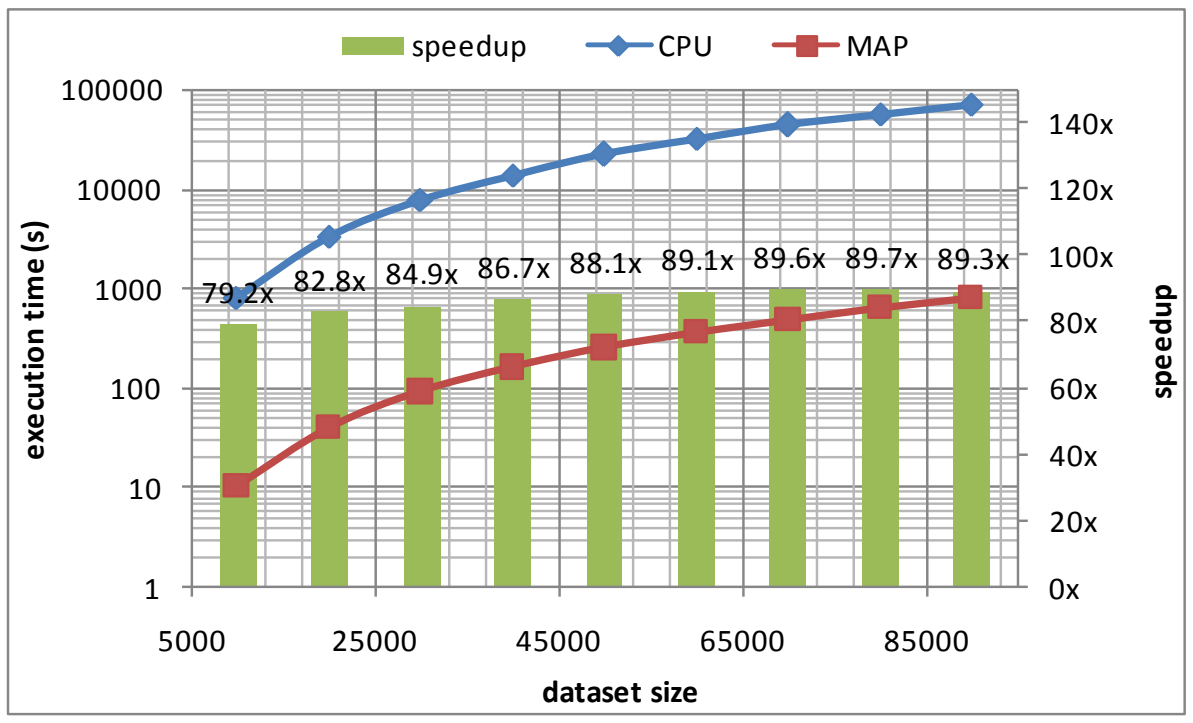

**Figure 5**: Performance results for the initial dual-MAP implementation of the TPACF algorithm.

The autocorrelation subroutine executes faster than the cross-correlation subroutine as it requires about half as many calculations as the cross-correlation subroutine. The exact execution time ratio is a function of the number of compute kernels used to implement the corresponding subroutines. Figure 6 shows the execution time of each of these subroutines as a function of the dataset size. Thus, for the dataset consisting of 10,000 points, the autocorrelation subroutine (running exclusively on MAP Series C processor) executes in 7.6 seconds (there are 101 calls to this subroutine), whereas the cross-correlation subroutine (running exclusively on MAP Series E processor) executes in 10.2 seconds (there are 100 calls to this subroutine), a 2.6 seconds difference. For the dataset consisting of 90,000 points, the difference is 224 seconds. In other words, the MAP Series C processor is idle about 18% of the time while the MAP

Series E processor is fully utilized. As the dataset size increases, the relative idle time for the MAP Series C processor stays the same (around 18%) while the absolute idle time grows quadratically.

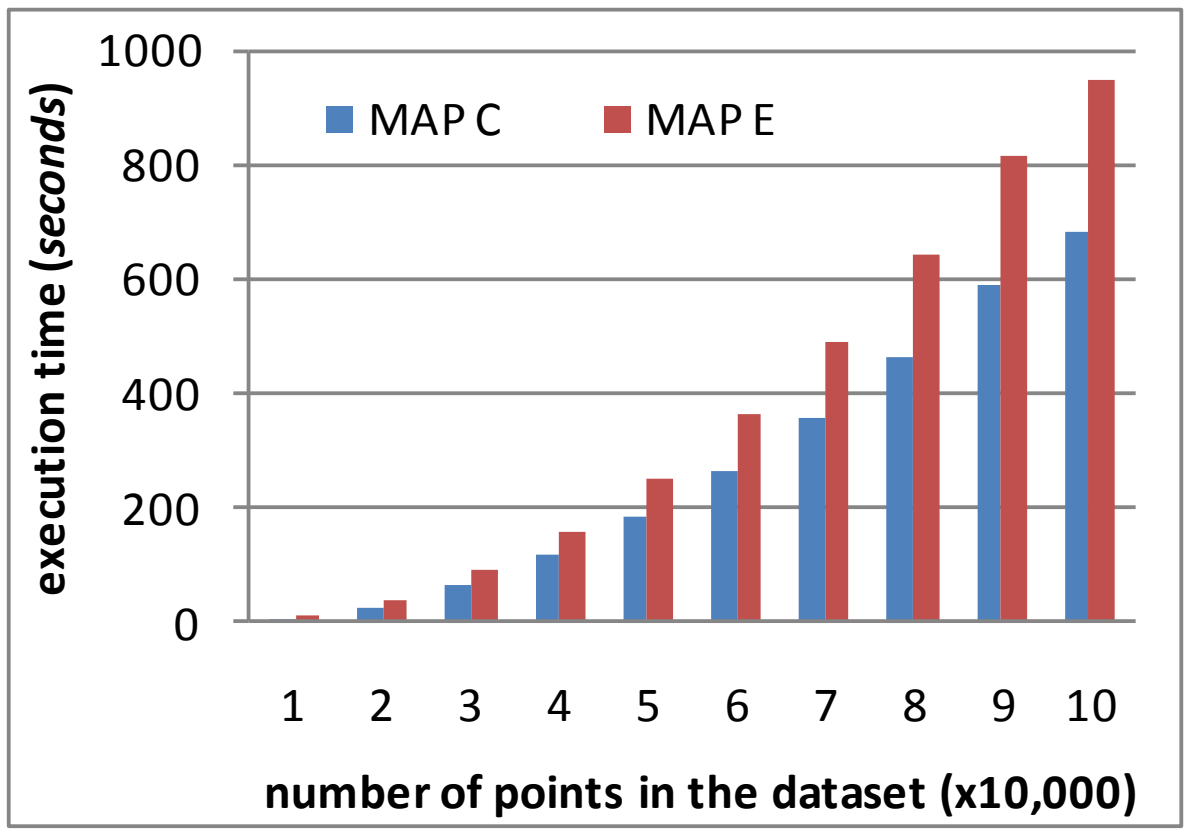

**Figure 6**: Execution time for the individual kernels (MAP processors) as a function of the dataset size.

Thus, while this implementation is simple and straightforward, it fails to fully utilize the available resources, since the MAP Series C processor remains idle about 18% of the time.

## 3.2. Theoretical performance analysis

Consider a simplified example in which only a single compute engine is implemented on each MAP processor and no data transfer or other overheads are taken into account. The compute time (measured as the number of steps necessary to execute the calculations in our general application, each step meaning a complete set of calculations for a pair of points) for a single pass over a random data file for this simplified kernel implementation is shown in Figure 7. In this simple example, the MAP Series C processor is idle about 50% of the time.

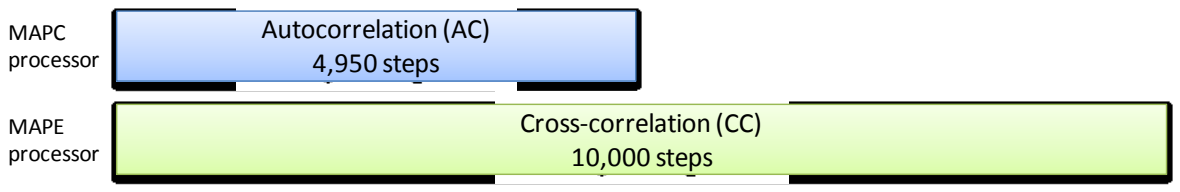

**Figure 7**: Number of compute steps necessary to process one random data file consisting of 100 points. Autocorrelation between the points in the random data file requires (100*100-1)/2=4,950 steps whereas the cross-correlation between the observed data and random data requires 100*100=10,000 steps.

Now consider an example in which the same FPGA kernel is used, but the observed data and random data sets are divided into 3 equally sized partitions and two MAP processors are scheduled to work in parallel on the same data file rather than on two data files at once. In this case, the MAP Series C processor will be first invoked in the autocorrelation mode to process the first partition of the random data file, while the MAP Series E processor will be invoked to compute the cross-correlation between the first and second partition of the random data file (Figure 8, blue bars). Once the MAP Series C processor is done with its first job assignment, it will start executing the cross-correlation between the first and the third partition of the random data file, and so on. Once all the segments belonging to the random data set are processed, the autocorrelation computation for this set is complete and the MAP processors can be allocated to compute the cross-correlation between the observed and random data sets (Figure 8, green bars). As a result, the MAP Series C processor is idle only about 6% of the time as compared to the MAP Series E processor and the overall execution time is equal to 76% of the execution time of the implementation from Figure 7.

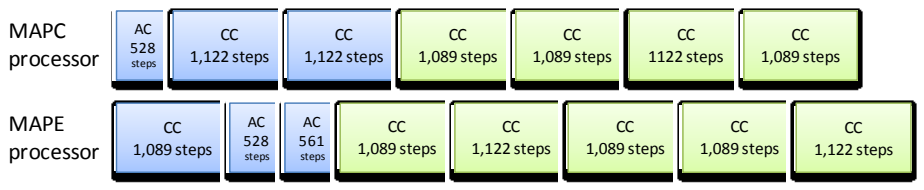

**Figure 8**: Number of compute steps necessary to execute the same problem as shown in Figure 7 when compute tasks are reallocated per MAP processor rather than per data file.

Clearly, this approach demonstrates a potential for an overall performance improvement as compared to the original, straightforward implementation. The exact amount of performance improvement, of course depends on a number of factors, such as the number of compute engines implemented per chip, amount of overhead due to the data transfer, pipelined loop depth, etc.

## 3.3. Load-balanced implementation

Implementing this load-balancing technique in practice is more complex: we replace calls to the autocorrelationMAPC and the cross-correlationMAPE subroutines with a call to a job scheduler that partitions the data sets into appropriately sized segments and invokes the original MAP-based subroutines to work on individual segments (autocorrelation) or segment pairs (cross-correlation) until all of them are processed. Partial results from each call to the MAP-based subroutines are accumulated and merged at the end of the execution.

We implemented the job scheduler as a simple loop that iterates on all the segments/segment pairs (tasks) to be processed and schedules each such task for the execution on the first available MAP processor. Individual tasks are scheduled as pthreads. Each dataset is divided into 5 equal-sized segments, although a higher granularity would be desirable for larger datasets.

Figure 9 contains the performance results for this implementation. We observe that for the smallest dataset used, the penalty for invoking the MAP processor-based code is rather high and the performance of this implementation is only about 46.5 times better than the performance of a single 2.8 GHz Intel Xeon chip-based implementation, while in our previous implementation (Figure 5), the performance ratio was 79.2x. However, with increasing dataset size, we start to observe the performance improvements beyond those achieved by our previous implementation. Thus, for the largest dataset processed, the speedup obtained with this load-balanced implementation is **96.2x** as compared to the speedup of only 89.3x achieved with our first implementation, a 9% overall application performance improvement. This corresponds to 8 GFLOPs of sustained floating-point performance as compared to 7.4 GFLOPS in our original dual-MAP implementation, or 83 MFLOPs in the reference C implementation.

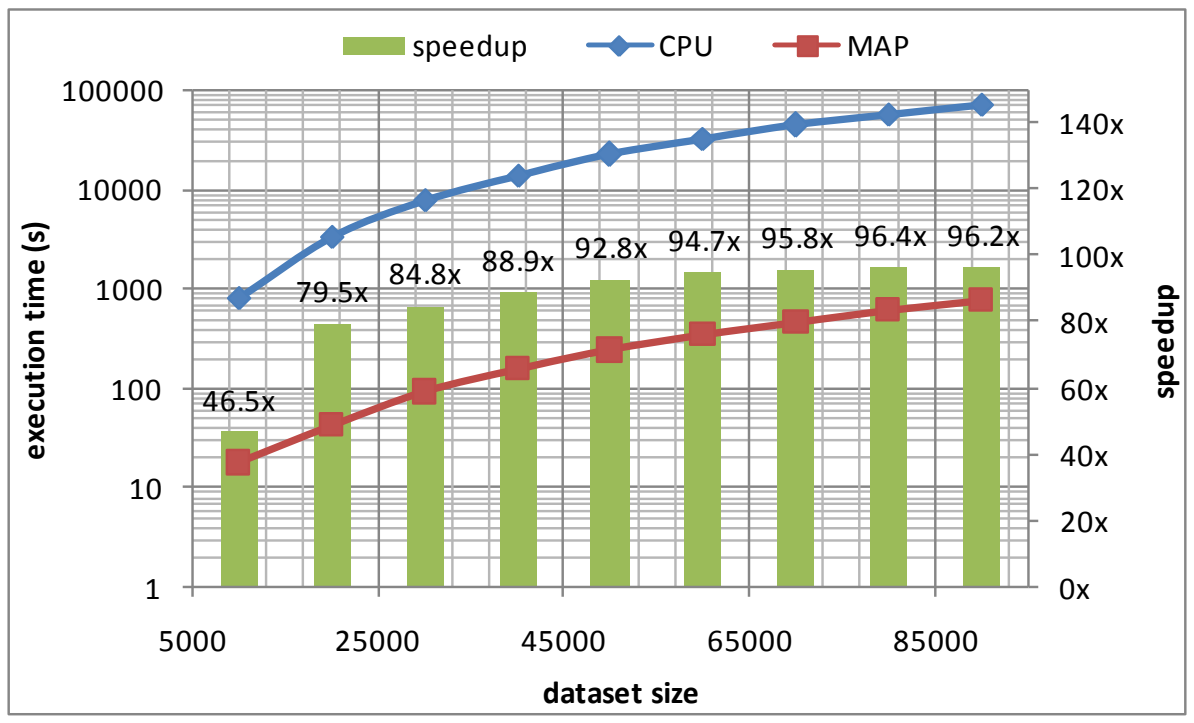

**Figure 9**: Performance results of the load-balanced dual-MAP implementation of the TPACF algorithm.

Figure 10, in a manner similar to Figure 6, shows the amount of time spent by each of the MAP processors executing the calculations as a function of the dataset size. In contrast to Figure 6, however, none of the MAP processors were used to exclusively execute the autocorrelation or cross-correlation subroutines. We observe that for a dataset consisting of 90,000 data points, the MAP Series C processor was invoked 1,808 times, while the MAP Series E processor was invoked 2,207 times. We also observe that the MAP Series C processor spent about 6 seconds longer performing the calculations than the MAP Series E processor. Thus, the MAP Series E processor is idle less than 1% of the overall execution time as compared to the 18% idle time of our initial implementation as discussed earlier. Clearly, load-balancing the work across multiple MAP processors produced a much better hardware utilization that also resulted in a better overall performance.

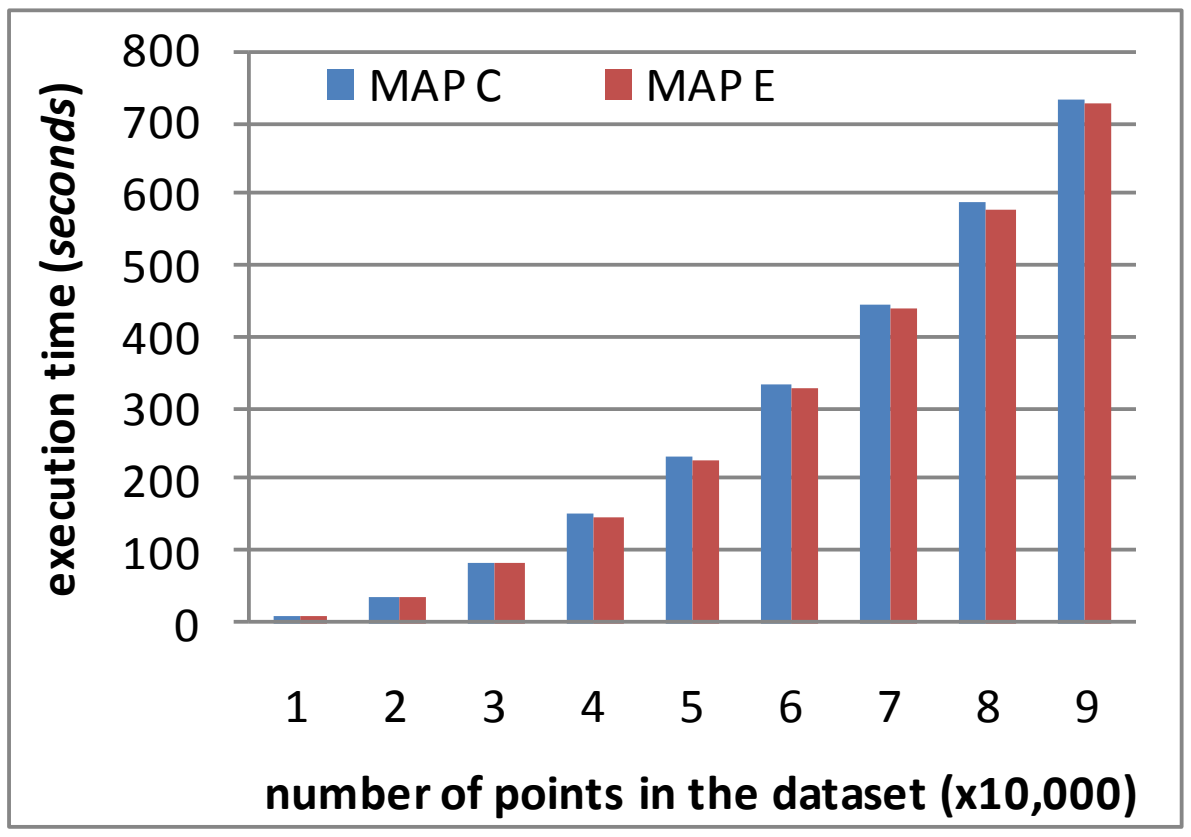

**Figure 10**: Execution time for the individual MAP processors as a function of the dataset size.

## 4. Conclusions

In this case study, we investigated a workload balancing strategy in which the execution of two kernels was serialized, while the execution of each individual kernel was parallelized by spreading the workload between two MAP processors. This is in contrast to our original implementation where both kernels were executed in parallel on two MAP processors. While the implementation of this approach is somewhat more involved, the benefits in terms of the overall application performance are substantial: the execution time of the application was reduced by 9% while maximizing the use of hardware resources.

## 5. Acknowledgements

This work was funded by the National Science Foundation grant SCI 05-25308 and by NASA grant NNG06GH15G. We would like to thank David Caliga, Dan Poznanovic, and Dr. Jeff Hammes, all from SRC Computers Inc., for their help and support with SRC-6 system.